\documentclass[doublecol]{epl2}
\usepackage{amssymb,amsfonts}
\usepackage{relsize}
\usepackage{graphicx}
\usepackage{dcolumn}
\usepackage{amsmath}
\usepackage{bm}
\usepackage{epsfig}

\newcommand{\be}{\begin{equation}}
\newcommand{\ee}{\end{equation}}

\newcommand{\mediaT}[1]{\left\langle #1 \right\rangle}

\title{Fluctuations of motifs and non self-averaging in complex networks\\
A self- \textit{vs} non-self-averaging phase transition scenario}
\shorttitle{A self- \textit{vs} non-self-averaging phase transition scenario in complex networks}
\author{M. Ostilli \inst{1}}
\institute{ \inst{1}
Cooperative Association for Internet Data Analysis, 
San Diego Supercomputer Center, UCSD, San Diego, CA\\
}


\abstract{
Complex networks have been mostly characterized from the point of view of the degree distribution of their nodes and a few other motifs (or modules), with
a special attention to triangles and cliques. The most exotic phenomena have been observed when the exponent $\gamma$ of the associated
power law degree-distribution is sufficiently small. In particular, a zero percolation threshold takes place for $\gamma<3$, and an anomalous critical behavior sets in for
$\gamma<5$. In this Letter we prove that in sparse scale-free networks characterized by a cut-off scaling with the sistem size $N$, 
relative fluctuations are actually never negligible: given a motif $\Gamma$, we analyze the relative fluctuations
$R_{\Gamma}$ of the associated density of $\Gamma$, and we show that there exists an interval in $\gamma$, $[\gamma_1,\gamma_2]$, 
where $R_{\Gamma}$ does not go to zero in the thermodynamic limit,
where $\gamma_1\approx k_{\mathrm{min}}$ and $\gamma_2\approx 2 k_{\mathrm{max}}$, $k_{\mathrm{min}}$ and $k_{\mathrm{max}}$ being the smallest and the largest degree of $\Gamma$, respectively.
Remarkably, in $(\gamma_1,\gamma_2)$ $R_{\Gamma}$ diverges, implying the instability of $\Gamma$ to small perturbations. 
}

\begin{document}

\maketitle

\email{massimo.ostilli@gmail.com}

\section{Introduction}
In the last decade, several complex networks models have been proposed 
to explain and reproduce the widespread presence of real-world networks \cite{Bollobas,Barabasi,Dorog,NewmanBook}.  
{Observed real-world networks are the output of certain random processes. Therefore, 
a complex network model should reproduce not only the same observed averages
but also the same observed sample to sample fluctuations, if available.} 
If data about fluctuations are not
available, the model should remain maximally random around the observed averages \cite{ERG}
or make use of a minimal number of assumptions (a null model approach).

\begin{figure}[tbh]
{\includegraphics[height=1.5in]{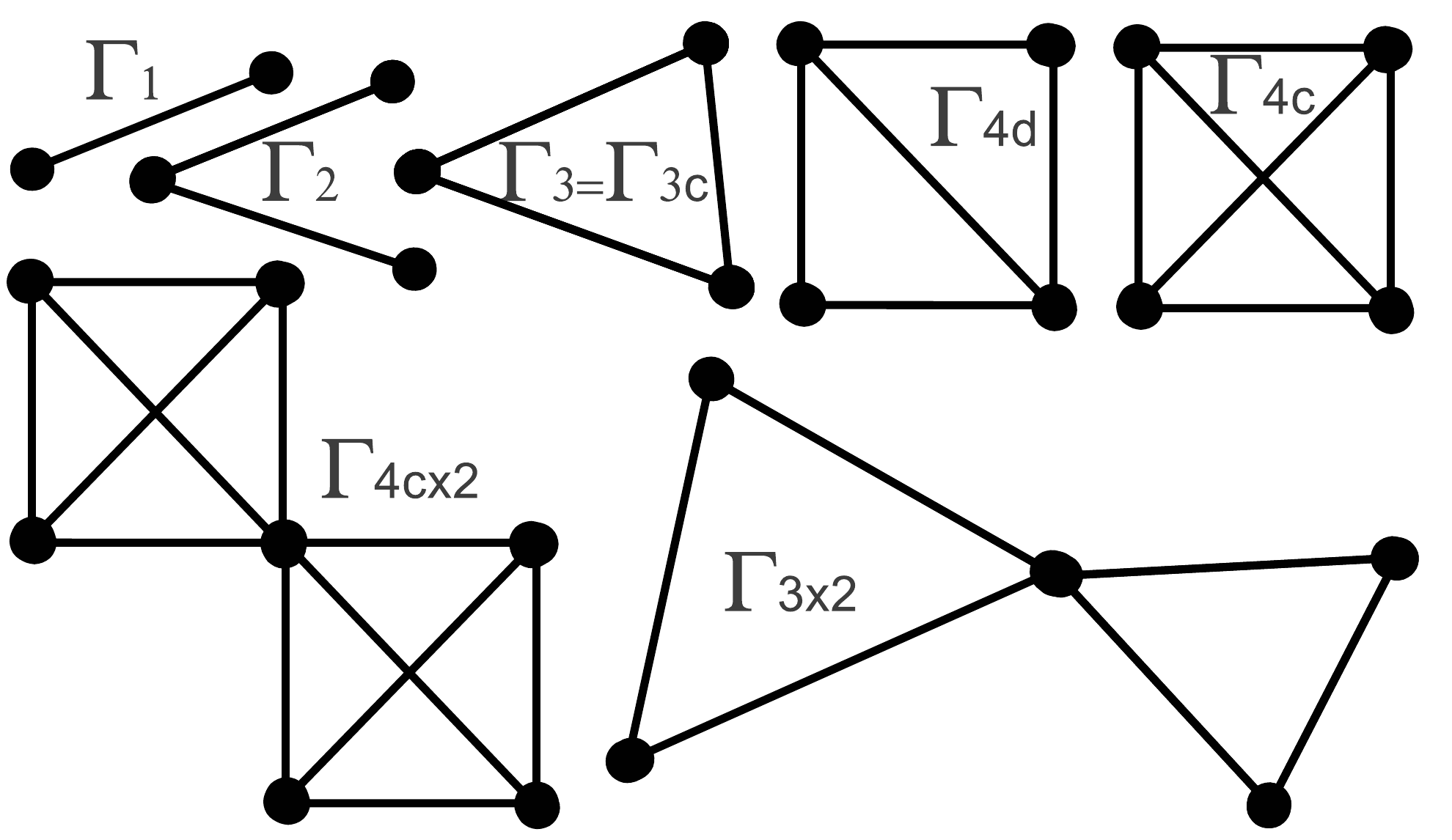}}
\caption{Examples of motifs. The labels in $\Gamma$ may specify the number of links of the motif;
for motifs made of $k$ fully connected nodes, $k$-cliques, we also use the symbol $\Gamma_{kc}$; $d$ in $\Gamma_{4d}$ stands for the presence of a diagonal;
$\Gamma_{kc\times 2}$ stands for two $k$-cliques sharing a common node.  
\label{fig1}
}
\end{figure}

In  all branches of physics, 
fluctuations have played a crucial role in the understanding of 
the  underlying phenomena.  It is not  excessive to say that
any observable in physics does not have any objective meaning without  
the evaluation of its fluctuations. Observables in networks are not an
exception. Particularly important are the motifs, also known as
modules, patterns, or communities~\footnote{
Here the term ``community'' is meant in a broad sense. In general a community is
a motif, but not \textit{vice-versa} \cite{Santo}.}
(Fig. \ref{fig1}), \textit{i.e.}, sub-graphs on which the functionality of the network largely depends \cite{Santo,Milo,BaraProc,BaraBio}.
In this Letter we provide  a first systematic
analysis of the  fluctuations of the density of  motifs  in an analytically treatable  model of networks, the hidden variable model 
\cite{CaldaHidden,BogunaHidden,NewmanHidden,SatorrasHidden,BogunaHidden2},
and we show  that these fluctuations, in certain  regions of the model
parameters,  are far from being  negligible.  
In fact, we discover the existence
of a phase transition  scenario with  regions  
where  relative   fluctuations  are negligible,  separated from  regions where 
the  relative fluctuations diverge.  
The  practical consequences  of this analysis  are dramatic.
A large class of real networks can be mapped toward suitable hidden variable models.
For each real network realization such a mapping requires measuring the hidden variables $\{h_i\}$ associated to each node.
The sequence $\{h_i\}$, however, in the process that produces the real network, can vary, and different real network realizations
in general are characterized by different sequences $\{h_i\}$. It is this variability of the sequence $\{h_i\}$ that generates extremely large fluctuations. 
In particular, two large network realizations 
will present two totally different community structures.
More in general, large fluctuations translate in a effective instability of the substructures composing the network to small perturbations. 
For the same reason, in simulating  complex  networks,  if we are in the high fluctuation region,
in order to have a fair evaluation of the average of the density of a motif
we need to make use of a very demanding statistics.
   
Complex networks studies have been mostly focused on the analysis of local averages and variances,
but without a systematic study of fluctuations. 
Analysis of correlations in complex networks have been in fact mainly confined
to the degree of adjacent nodes \cite{Dorog,NewmanBook,BogunaHidden}, or to the presence of loops and cliques \cite{Bianconi,Bianconi1,Bianconi2}
which are a manifestation of correlations. 
Particular attention has been paid to the ``configuration model'' \cite{Bollobas,Dorog},
\textit{i.e.}, the ``uncorrelated'' network model''~\footnote{
The expression ``uncorrelated network model'' might be misleading, but we keep it using for historical reasons.
A network is said to be ``uncorrelated'' if the degrees $k$ and $k'$ at the ends of a link are independent random variables.
However, a lack of degree-degree correlations does not imply the absence of other correlations.
}
originating from all possible graphs constrained to
satisfy a given degree distribution exactly (hard version) \cite{Bollobas,Dorog,Samukhin,Bianconi3}, or in average (soft version) \cite{Goh,BogunaHidden,CaldaHidden,NewmanHidden}. 
It is well known that,
in the thermodynamic limit, in the configuration model the presence of loops of finite length is negligible, 
provided the exponent $\gamma$ of the associated power law degree-distribution $P(k)\sim k^{-\gamma}$ is sufficiently large \cite{Dorog,Bianconi}.
In turn, this has led perhaps to the erroneous conclusion that, in any synthetic or real network, fluctuations
are always negligible for sufficiently large $\gamma$.
In fact, exotic phenomena are believed to occur only when $\gamma<5$,
where an anomalous critical behavior sets in, especially when $\gamma<3$,
where a zero percolation threshold takes place due to a divergent second moment of $P(k)$ \cite{Review}. 
In particular, fluctuations of motifs 
are assumed to be negligible for $\gamma$ enough large.
In this Letter, by using the framework of hidden-variable models, 
we prove that in sparse networks ($\gamma>2$) characterized by a cut-off scaling with the system size $N$, 
fluctuations are actually never negligible: given a motif $\Gamma$, we analyze the relative fluctuations
$R_{\Gamma}$ of the density of $\Gamma$, and we show that there exists an interval $(\gamma_1,\gamma_2)$ where $R_{\Gamma}$ diverges in the thermodynamic limit,
where $\gamma_1\approx k_{\mathrm{min}}$ and $\gamma_2\approx 2 k_{\mathrm{max}}$, $k_{\mathrm{min}}$ and $k_{\mathrm{max}}$ being the smallest and the largest degree of $\Gamma$.
As a consequence, in $(\gamma_1,\gamma_2)$, measuring the density of $\Gamma$ in simulations is a hard problem, and $\Gamma$ is unstable to small perturbations,
a fact that in turn provides a key to understand the stability/instability of communities \cite{Santo}.

\section{The hidden-variable scheme. The choice of the cut-off}
\begin{figure}[tbh]
{\includegraphics[width=3.2in]{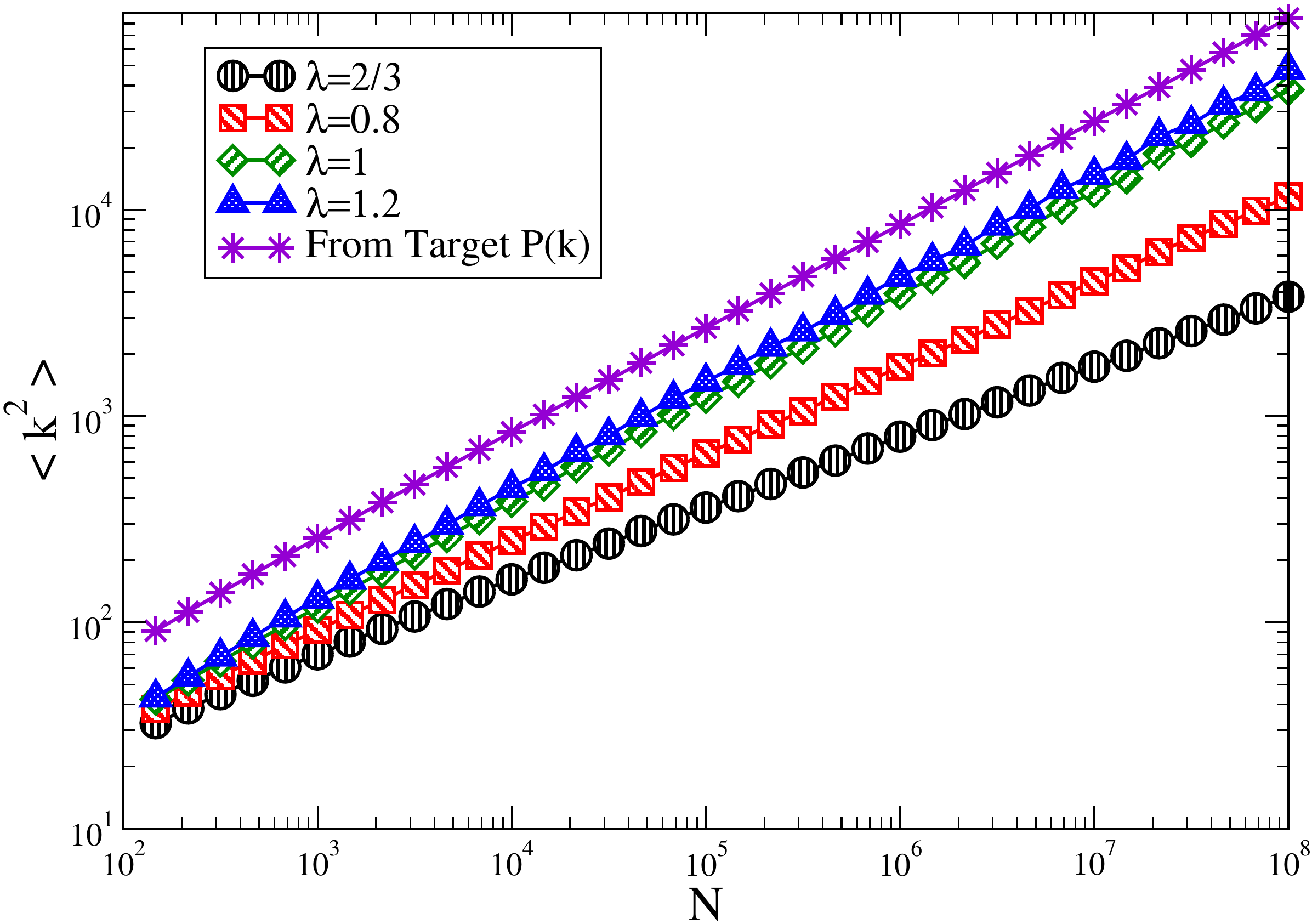}}
\caption{Behavior of $\mediaT{k^2}$ \textit{vs} the system size $N$ for $\gamma=2.5$. The upper plot corresponds to the target degree distribution which is a
power law $P(k)=ak^{-\gamma}$ with $k\leq N$,
whereas the other plots correspond to the hidden variable model (\ref{CM})-(\ref{rho}) with different choices of the cut-off $h_{\mathrm{max}}=N^\lambda$: 
$\lambda=1/(\gamma-1)=2/3$ (corresponding to the ``natural cut-off''), $\lambda=0.8$, $\lambda=1$, and $\lambda=1.2$. For higher values of $\lambda$,
the plots saturate to a curve that, on this scale, is indistinguishable from the case $\lambda=1.2$. 
The plots of the hidden-variable model have been calculated by numerical evaluation of the involved integrals:
$\mediaT{k^2}\simeq N^2\int_{h_{\mathrm{min}}}^{h_{\mathrm{max}}}dhdh'dh''\rho(h)\rho(h')\rho(h'')p(h,h')p(h',h'')$ (see below for a more detailed analysis
of these techniques).
\label{figs}
}
\end{figure}

Given $N$ nodes, hidden variable models are defined in the following way:
\textit{i)}
to each node we associate a hidden variable $h$ drawn from
a given probability density function (PDF) $\rho(h)$; 
\textit{ii)} between any pair of nodes, we assign, or not assign, a link, according
to a given probability $p(h,h')$, where $h$ and $h'$ are the hidden variables associated to the two nodes.
The probability $p(h,h')$ can be any function of the $h$'s, the only requirement being that $0 \leq p(h,h')\leq 1$.
It has been shown that, when $p(h,h')$ has the following form (or similar generalizations)~
\begin{eqnarray}
\label{CM}
&& p(h,h')= \left(1+\frac{k_s^2}{hh'}\right)^{-1}, \quad
k_s=\sqrt{N\bar{k}},
\end{eqnarray}
and $\bar{k}$ is the wanted average degree, for large $N$, the actual degree $k$ of the nodes of the network realized with the above scheme
are distributed according to $\rho$ with actual average degree equal to $\bar{k}$.
In particular, if we choose the following PDF having support in $[h_{\mathrm{min}},h_{\mathrm{max}}]$
\begin{eqnarray}
\label{rho}
\rho(h)=a~h^{-\gamma}, \quad h_{\mathrm{max}} \geq h \geq h_{\mathrm{min}}>0,
\end{eqnarray}
with $\gamma>2$,
the degree-distribution of the resulting network will be a power law with exponent $\gamma$ and, 
for $N$ sufficiently large, the normalization constant $a$ and the so called structural cut-off $k_s$ are
$a=(\gamma-1)/(h_{\mathrm{min}}^{1-\gamma})$, and $k_s=\sqrt{Nh_{\mathrm{min}}(\gamma-1)/(\gamma-2)}$.
When $h_{\mathrm{max}}\ll k_s$, correlations of the generated network are negligible, and $p(h,h')\simeq hh'/k_s^2$,
while for $h_{\mathrm{max}}\gg k_s$ correlations can be important. 
{The choice of the cut-off $h_{\mathrm{max}}$ is in principle arbitrary. 
However, most of real-world networks show that the maximal degree scales according to the so called natural cut-off: $k_{\mathrm{max}}\sim h_{\mathrm{nc}}=N^{1/(\gamma-1)}$. 
As a consequence, in several models of complex networks it was assumed the choice $h_{\mathrm{max}}=h_{\mathrm{nc}}$,
justified as empirical. We think however that such an approach is wrong: the fact that in most of the real-world networks  
$k_{\mathrm{max}}\sim h_{\mathrm{nc}}$ is due to a probabilistic effect, is not due to a rigid upper bound $k\leq h_{\mathrm{nc}}$. 
In fact, by using order-statistics one finds
that by drawing $N$ degree values from a power law with exponent $\gamma$,
the highest degree in average scales just as $\mediaT{k_{\mathrm{max}}}\sim N^{1/(\gamma-1)}$ \cite{DorogCut,BogunaCut}.
More precisely, it is possible to prove that the PDF for the rescaled random variable $k_{\mathrm{max}}/N^{1/(\gamma-1)}$ 
is also a power law with exponent $\gamma$ \cite{Remco}. 
Power law distributions always lead to important fluctuations. 
It is then clear that empirical observations of $k_{\mathrm{max}}$ must be taken with care: 
$k_{\mathrm{max}}$ is not a self-averaging variable and samples in which $k_{\mathrm{max}}\gg N^{1/(\gamma-1)}$,
even if extremely rare, do exist and, as we shall see, have dramatic effects on the fluctuations of motifs. 
We stress that here we follow a null-model approach. We do not claim that the degree of all real networks must have a cut-off scaling
with $N$; there might be of course many other possible scalings whose value depend on the details of the system (related to physical, biological, or economical constraints). 
However, if the only information that we have from a given real network of size $N$ is that \textit{i)} the degree obeys a power-law distribution with exponent $\gamma$, 
and \textit{ii)}  highest degrees scale in average as $N^{1/(\gamma-1)}$,    
forcing the model to have a specific cut-off other than $N$ would introduce a bias. In fact, order statistics tells us that a lower cut-off would
produce $\mediaT{k_{\mathrm{max}}}\ll N^{1/(\gamma-1)}$.
This observation leads us to choose $h_{\mathrm{max}}\sim N$ for the hidden-variable scheme (\ref{CM})-(\ref{rho}). 
More precisely, if we consider as target degree distribution a power-law $P(k)\propto k^{-\gamma}$ with finite support $k\leq N$,  
in order to reproduce its characteristics from the hidden-variable model,
we need to use a cut-off $h_{\mathrm{max}}=\mathop{O}(N^\lambda)$ with $\lambda\geq 1$. In fact, any other choice implies a difference
in the scaling of the moments $\mediaT{k^n}$ between the hidden variable model (\ref{CM})-(\ref{rho}) and the target distribution $P(k)$.}
Fig. \ref{figs} shows this for the second moment. Similar plots hold for higher moments. In conclusion, the minimal cut-off of the model (\ref{CM})-(\ref{rho})
able to reproduce the correct scaling of all the moments of the target degree distribution $P(k)$ is just $h_{\mathrm{max}}=\mathop{O}(N^\lambda)$ with $\lambda=1$. 
In this paper we set therefore $h_{\mathrm{max}}=N$. 
We stress that with this choice highest degrees will be still order $N^{1/(\gamma-1)}$, but only on average.  

\section{Fluctuations of Motifs}
Given the parameters $N$, $h_{\mathrm{min}}$, $\bar{k}$, and $\gamma$,
the above hidden-variables scheme produces an ensemble of networks which, in terms of a few characteristics, like statistics
of the degree and motifs, are in part representative of many real-world networks with those given parameters. 
In the following we will indicate the ensemble averages with the bracket symbol $\mediaT{\cdot}$.
The averages are built by following the above steps \textit{(i)} and \textit{(ii)} of the hidden-variables scheme.
Notice that each time we generate a network realization, we need to draw
$N$ hidden variables from the PDF $\rho(h)$, and $N(N-1)/2$ numbers to sample $p(h,h')$.
In terms of the adjacency matrix $a_{i,i}$, taking value 0 or 1 for the presence or not of a link between nodes $i$ and $j$,
steps \textit{(i)} and \textit{(ii)} give
\begin{eqnarray}
\label{link0}
\mediaT{a_{i,j}}=\int dh_idh_j\rho(h_i)\rho(h_j)p(h_i,h_j).
\end{eqnarray}

We will indicate by $n_\mathsmaller{\Gamma}$ the density of the motif $\Gamma$ in a network realization. 
As is known \cite{BogunaHidden,BogunaC}, for $\gamma>2$, the hidden variable model defined through Eqs. (\ref{CM})-(\ref{rho}) leads to a small clustering coefficient
$C=\mediaT{n_\mathsmaller{\Gamma_3}}/\mediaT{3n_\mathsmaller{\Gamma_2}}$.
For example, for $\gamma\gg 3$ we have $\mediaT{n_\mathsmaller{\Gamma_2}}=\mathop{O}(1)$, while
$\mediaT{n_\mathsmaller{\Gamma_3}}=\mathop{O}(1/N)$. 
More in general, the more the motif is clustered, the smaller is its density.
Yet, for finite $N$, and for any motif $\Gamma$, clustered or not, 
by tuning the parameters $h_{\mathrm{min}}$, $\bar{k}$ and $\gamma$, one can set, within some freedom, 
a desired value of $\mediaT{n_\mathsmaller{\Gamma}}$.
However, as we shall see, the sample-to-sample fluctuations of $n_\mathsmaller{\Gamma}$ 
can be unexpectedly large. 
Fluctuations of $n_\mathsmaller{\Gamma}$ must be compared with the corresponding average of $n_\mathsmaller{\Gamma}$,
therefore we are going to analyze the following standard ratio
\begin{eqnarray}
\label{R}
R_\mathsmaller{\Gamma}=\frac{\mediaT{{n_\mathsmaller{\Gamma}}^2}-\mediaT{n_\mathsmaller{\Gamma}}^2}{\mediaT{n_\mathsmaller{\Gamma}}^2}.
\end{eqnarray}
In general $R_\mathsmaller{\Gamma}$ will strongly depends on $\Gamma$, $N$ and $\gamma$.
When $\lim_{N\to\infty} R_\mathsmaller{\Gamma}=0$ the network is said to be self-averaging with respect to the motif density $n_\mathsmaller{\Gamma}$.
In practical terms, when this occurs, even one single sample is enough to get by simulations an accurate estimation of the average $\mediaT{n_\mathsmaller{\Gamma}}$,
provided $N$ is large enough. The behavior of $R_\mathsmaller{\Gamma}$ with respect to the network size $N$ is therefore of crucial importance:
if the network is not self-averaging with respect to some motif $\Gamma$, the number of samples necessary to get 
a good estimation of $\mediaT{n_\mathsmaller{\Gamma}}$ in simulations will have to grow with $N$
or, from another perspective, it is hard to generate only those samples whose density is close to a target value, 
and a kind of hard searching problem emerges. This aspect is in fact connected 
with spin-glass and NP-complete problems; we will see in fact that $R_\mathsmaller{\Gamma}$
can be read as a susceptibility of a homogeneous system.\\

\section{Analysis of $R_\mathsmaller{\Gamma}$}
Given a motif $\Gamma$, the density of  $\Gamma$ in a graph realization is
\begin{eqnarray}
\label{nM}
n_\mathsmaller{\Gamma}=\frac{c}{N}\sum_i k_{\mathsmaller{\Gamma}}(i),
\end{eqnarray}
where $k_{\mathsmaller{\Gamma}}(i)$ counts the number of motifs $\Gamma$ passing through the node $i$. 
The coefficient $c$ depends on the definition
of the motif considered and serves to avoid over-counting when the motif is symmetric. For example,
if the motif $\Gamma$ is the triangle, we set $c=1/3$. If the motif is not symmetric, we can establish to count
only those motifs that pass through a specif node of $\Gamma$. For example, if $\Gamma$ is a triple (two consecutive links), 
we can set $c=1$, but a motif contributes only when the center of the triple coincides with $i$.
However, since we are interested only in the relative fluctuations $R_\mathsmaller{\Gamma}$, we do not need to specify it since $c$, as well as any constant, 
does not play any role for $R_\mathsmaller{\Gamma}$. Let us consider now the numerator of Eq. (\ref{R}). 
Note that the hidden variable scheme does not distinguish nodes, therefore we 
can make use of the fact that nodes are all statistically equivalent. 
By using this property, from Eq. (\ref{nM}) we get the following susceptibility
\begin{eqnarray}
\label{RN}
&& \mediaT{{n_\mathsmaller{\Gamma}}^2}-\mediaT{n_\mathsmaller{\Gamma}}^2=
\frac{c^2}{N}\left\{\mediaT{k^2_{\mathsmaller{\Gamma}}(i)}-\mediaT{k_{\mathsmaller{\Gamma}}(i)}^2\right\}\nonumber \\
&+& c^2\left\{\mediaT{k_{\mathsmaller{\Gamma}}(i)k_{\mathsmaller{\Gamma}}(j)}_{i\neq j}-\mediaT{k_{\mathsmaller{\Gamma}}(i)}^2\right\},
\end{eqnarray}
where $i$ and $j$ represent two arbitrary distinct indices.
In the rhs of Eq. (\ref{RN}) we have a self-term proportional to the motif variance, 
rescaled by the factor $1/N$,
and a mixed-term that accounts for correlations between two motifs centered at two different nodes.
Note that, in general, the self-term, despite appears to be order $1/N$, cannot be
neglected. In fact, due to exact cancellations in the mixed term, the mixed- and self-terms give contributions of the same order of magnitude.  

For what follows,
we find it convenient to introduce another symbol for the averages with respect to the PDF $\rho(h)$ given by Eq. (\ref{rho}):
if $f(\cdot)$ is any function of the hidden variables $h_1,\ldots,h_N$ we define
\begin{eqnarray}
\label{rhoav}
\left[f\right]=\int \prod_{i=1}^N dh_i\rho(h_i)f(\cdot).
\end{eqnarray}
In particular, from Eq. (\ref{link0}) we have
$\mediaT{a_{i,j}}=\left[p(h_i,h_j)\right]$.
For $\gamma>2$, $k_s\sim N^{1/2}$, therefore Eqs. (\ref{CM}) and (\ref{rho}) imply
\begin{eqnarray}
\label{rhoav2}
\mediaT{a_{i,j}}=\left[p(h_i,h_j)\right]=\mathop{O}\left(N^{-1}\right).
\end{eqnarray}
Next we analyze Eq. (\ref{RN}) in a few crucial motifs.

\begin{figure}[tbh]
{\includegraphics[width=1.5in]{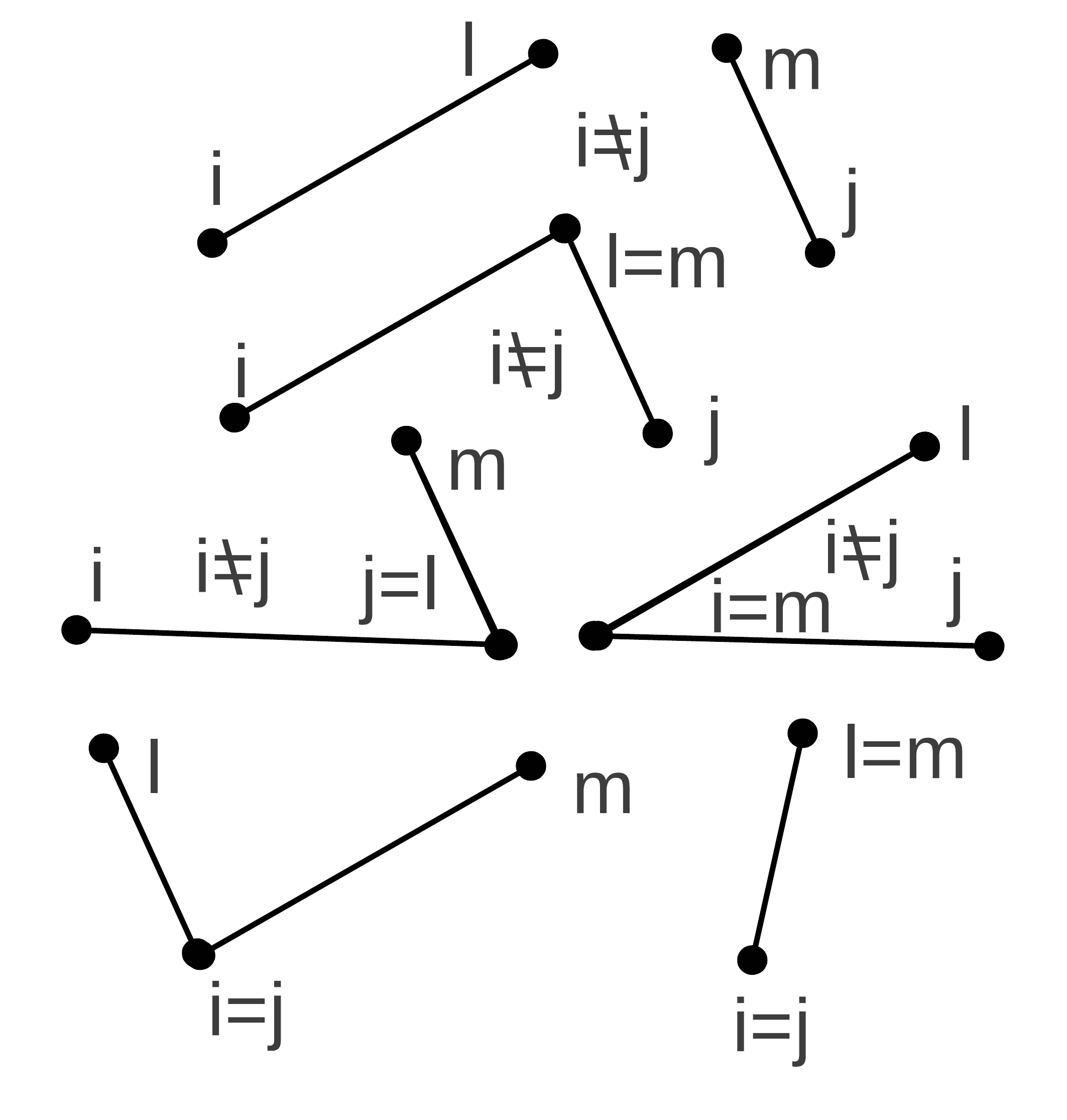}}
\caption{Contributions to Eqs. (\ref{link3}) (lower connected motifs) and (\ref{link4}) (upper disconnected motifs and the
3 connected motifs located in the central part of the figure). 
Nodes $i$ and $j$ are to be kept fixed, while the others can vary,
provided the topology is kept fixed.
Contributions from disconnected motifs always cancel in $R_\mathsmaller{\Gamma}$.
\label{fig2}
}
\end{figure}

\textit{Link} $(\Gamma_{1})$.
If $\Gamma$ is the link $k_{\mathsmaller{\Gamma}}(i)$ coincides
with the standard definition of degree of the node $i$. 
For a given graph realization, corresponding to a given realization of the $h$'s, in terms of adjacency matrix we have
\begin{eqnarray}
\label{link}
k_{\mathsmaller{\Gamma_{1}}}(i)=\sum_{l\neq i}a_{i,l}.
\end{eqnarray}
By using Eq. (\ref{link0}), Eqs. (\ref{rhoav})-(\ref{link}), and the statistical equivalence of nodes, we have
\begin{eqnarray}
\label{link1}
\mediaT{k_{\mathsmaller{\Gamma_{1}}}(i)}=\sum_{l\neq i}\mediaT{a_{i,l}}=(N-1)\left[p(h_1,h_2)\right].
\end{eqnarray}
Let us now consider the product $k_{\mathsmaller{\Gamma_{1}}}(i)k_{\mathsmaller{\Gamma_{1}}}(j)$.
Notice that $a^2_{i,j}=a_{i,j}$.
We have to distinguish the cases $i=j$ and $i\neq j$, see Fig.~\ref{fig2}. 
For $i=j$ we have
\begin{eqnarray}
\label{link3}
&&\mediaT{k^2_{\mathsmaller{\Gamma_{1}}}(i)}
=(N-1)(N-2)\left[p(h_1,h_2)p(h_2,h_3)\right]
\nonumber\\ && + (N-1)\left[p(h_1,h_2)\right],
\end{eqnarray}
while for $i\neq j$ we have
\begin{eqnarray}
\label{link4}
&& \mediaT{k_{\mathsmaller{\Gamma_{1}}}(i)k_{\mathsmaller{\Gamma_{1}}}(j)}
=(N-2)(N-3)\left[p(h_1,h_2)\right]^2
\nonumber\\ && + 3(N-1)\left[p(h_1,h_2)p(h_2,h_3)\right],
\end{eqnarray}
where the factor 3 comes from the fact that two links emanating from nodes $i$ and $j$ can share a same node in 3 
topologically equivalent ways. On plugging Eqs. (\ref{link1})-(\ref{link4}) into Eq. (\ref{R}) via Eq. (\ref{RN}) and 
keeping only terms in $N^2$, which cancel exactly, and terms in $N$, we obtain
\begin{eqnarray}
\label{RL}
R_\mathsmaller{\Gamma_{1}}=\frac{4}{N}\frac{\left[p(h_1,h_2)p(h_2,h_3)\right]}{\left[p(h_1,h_2)\right]^2}-\frac{4}{N}+\frac{1}{N^2\left[p(h_1,h_2)\right]}.
\end{eqnarray}
Due to Eq. (\ref{rhoav2}) and its generalizations, for $\gamma>3$, each term present in the rhs of Eq. (\ref{RL}) is of order $1/N$, therefore
we have $R_\mathsmaller{\Gamma_{1}}=\mathop{O}(1/N)$ and the network is self-averaging with respect to the link density, 
while for $\gamma<3$ we have still self-averaging but $R_\mathsmaller{\Gamma_{1}}$ decays slower as $N^{2-\gamma}$. 
It is interesting however to observe the general behavior of $R_\mathsmaller{\Gamma_{1}}$ with respect to $\gamma$ for finite $N$.   
As a general rule, $\left[p(h_1,h_2)p(h_2,h_3)\ldots p(h_m,h_{m+1})\right]$ for large $\gamma$ tends to factorize
: $\left[p(h_1,h_2)p(h_2,h_3)\ldots p(h_m,h_{m+1})\right]\to \left[p(h_1,h_2)\right]\left[p(h_2,h_3)\right]\ldots \left[p(h_m,h_{m+1})\right]$.
Therefore, the first two terms in the rhs of Eq. (\ref{RL}) tend to cancel for large $\gamma$. However, the last term does not cancel
for large $\gamma$ (this issue will be discussed elsewhere). 

\textit{Diagrammatic calculus.}
From Eq. (\ref{RL}) we see that the main term is given by the ratio between a $\left[\cdot\right]$-average of two links sharing a common node, and the square of 
the $\left[\cdot\right]$-average of a single link,\textit{i.e.}, our motif. 
From this example is clear that a correspondence between formulas and diagrams can be established to avoid unnecessary simulations and to improve
our understanding about the main contributions to $R_\mathsmaller{\Gamma}$, especially those that can generate non self-averaging. 
In this sense, we find it convenient to make use of the compact notation $\left[\Gamma\right]$, where $\Gamma$ can be any motif.
For example, by referring to Fig. \ref{fig1}, we have $\left[\Gamma_1\right]=\left[p(h_1,h_2)\right]$, $\left[\Gamma_2\right]=\left[p(h_1,h_2)p(h_2,h_3)\right]$,
$\left[\Gamma_3\right]=\left[p(h_1,h_2)p(h_2,h_3)p(h_3,h_1)\right]$,
and so on. The role played by these $\left[\cdot\right]$-averages, is similar
to the role played by Green functions in statistical field theory. Moreover, since $R_\mathsmaller{\Gamma}$ is defined in terms
of connected correlation functions (\ref{RN}), we need to work only with Green functions of connected motifs, as the contributions
of disconnected motifs always cancel. 
Next, by using this diagrammatic tool, we evaluate $R_\mathsmaller{\Gamma}$ in the crucial case of $k$-cliques.
We first analyze the case $k=3$ in detail, and then we look at the general behavior $R_\mathsmaller{\Gamma_{kc}}$,
omitting contributions which are not essential here. Further details will be given elsewhere.

\textit{Triangle} $(\Gamma_{3})$. This is the simplest $k$-clique. We have 
\begin{eqnarray}
\label{RT}
R_\mathsmaller{\Gamma_{3}}=\frac{9}{N}\frac{\left[\Gamma_{3\times 2}\right]}{\left[\Gamma_{3}\right]^2}-\frac{9}{N}+\frac{14}{N^2}\frac{\left[\Gamma_{4d}\right]}{\left[\Gamma_{3}\right]^2}
+\frac{6}{N^3}\frac{1}{\left[\Gamma_{3}\right]}.
\end{eqnarray}
The factor 9 comes from: 8 ways to build $\Gamma_{3\times 2}$ from the mixed term with $i\neq j$, and
an extra contribution from the self-term $i=j$.
Similarly to the last term of Eq. (\ref{RL}),
the last two terms of Eq. (\ref{RT}) do not cancel for large $\gamma$.  

\textit{$k$-Clique} $(\Gamma_{kc})$. Given ${\Gamma_{kc}}$ ($(k-1)$ is the degree of each node), if $\Gamma_{kc\times 2}$ indicates the motif
in which two $k$-cliques $\Gamma_{kc}$ share a common node, we have  
\begin{eqnarray}
\label{Rclique}
R_\mathsmaller{\Gamma_{kc}}=\frac{b_k}{N}\frac{\left[\Gamma_{kc\times 2}\right]}{\left[\Gamma_{kc}\right]^2}-\frac{b_k}{N}+\mathop{O}\left(\frac{1}{N}\right),
\end{eqnarray}
where $b_k$ is a combinatorial term which depends only on $k$, and the last term is positive and plays a role similar to the last two terms of Eq. (\ref{RT}).

\begin{figure}[tbh]
{\includegraphics[width=3.2in]{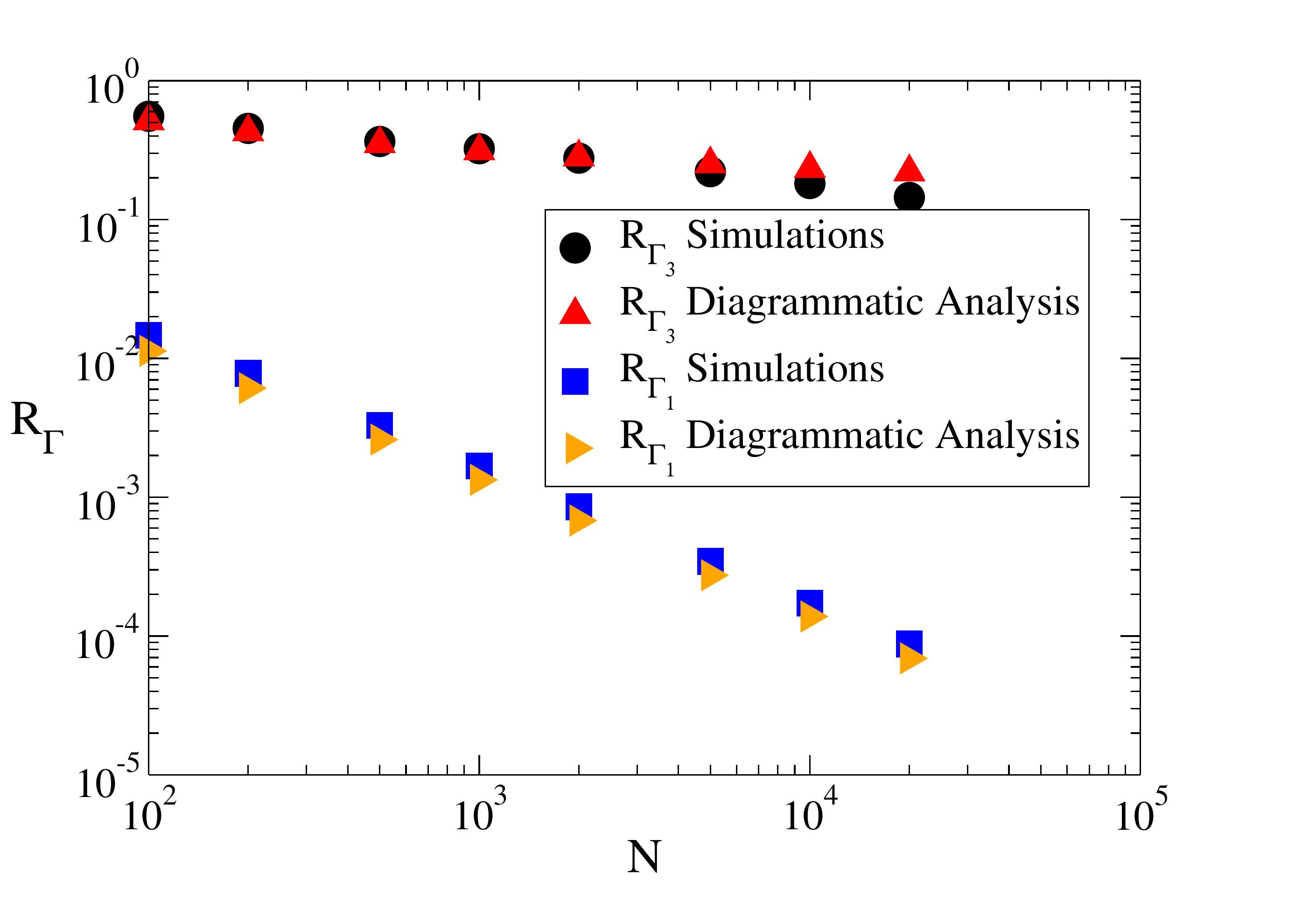}}
\caption{$R_\mathsmaller{\Gamma_1}$ and $R_\mathsmaller{\Gamma_3}$ as functions of $N$ for $\gamma=4.2$.
Simulations (circles and squares) made with $S=10^6$ samples for $N\in[10^2,2\cdot 10^3]$, $S=25\cdot 10^4$ for $N=5\cdot 10^3$, 
$S=4\cdot 10^4$ for $N=10^4$, and $S=10^4$ for $N=2\cdot 10^4$.  
Diagrammatic analysis (triangles) made by numerical integrations of Eqs. (\ref{RL}) and (\ref{RT}) by using $10^8$ points per integral.
\label{fig3}
}
\end{figure}

In Fig.~\ref{fig3} we show the behavior of $R_\mathsmaller{\Gamma_1}$ and $R_\mathsmaller{\Gamma_3}$ \text{vs} $N$ for $\gamma=4.2$, and show the matching 
simulations \textit{vs} diagrammatic analysis. Notice that the theory (see next paragraph) predicts $R_\mathsmaller{\Gamma_3}\to 0$ for $\gamma>4$,
and $R_\mathsmaller{\Gamma_3}\to \infty$ for $2.5<\gamma<4$,
however $\gamma=4.2$ is quite close to $4$ so that $R_\mathsmaller{\Gamma_3}$ decays very slowly with $N$.

\begin{figure}[tbh]
{\includegraphics[height=1.65in]{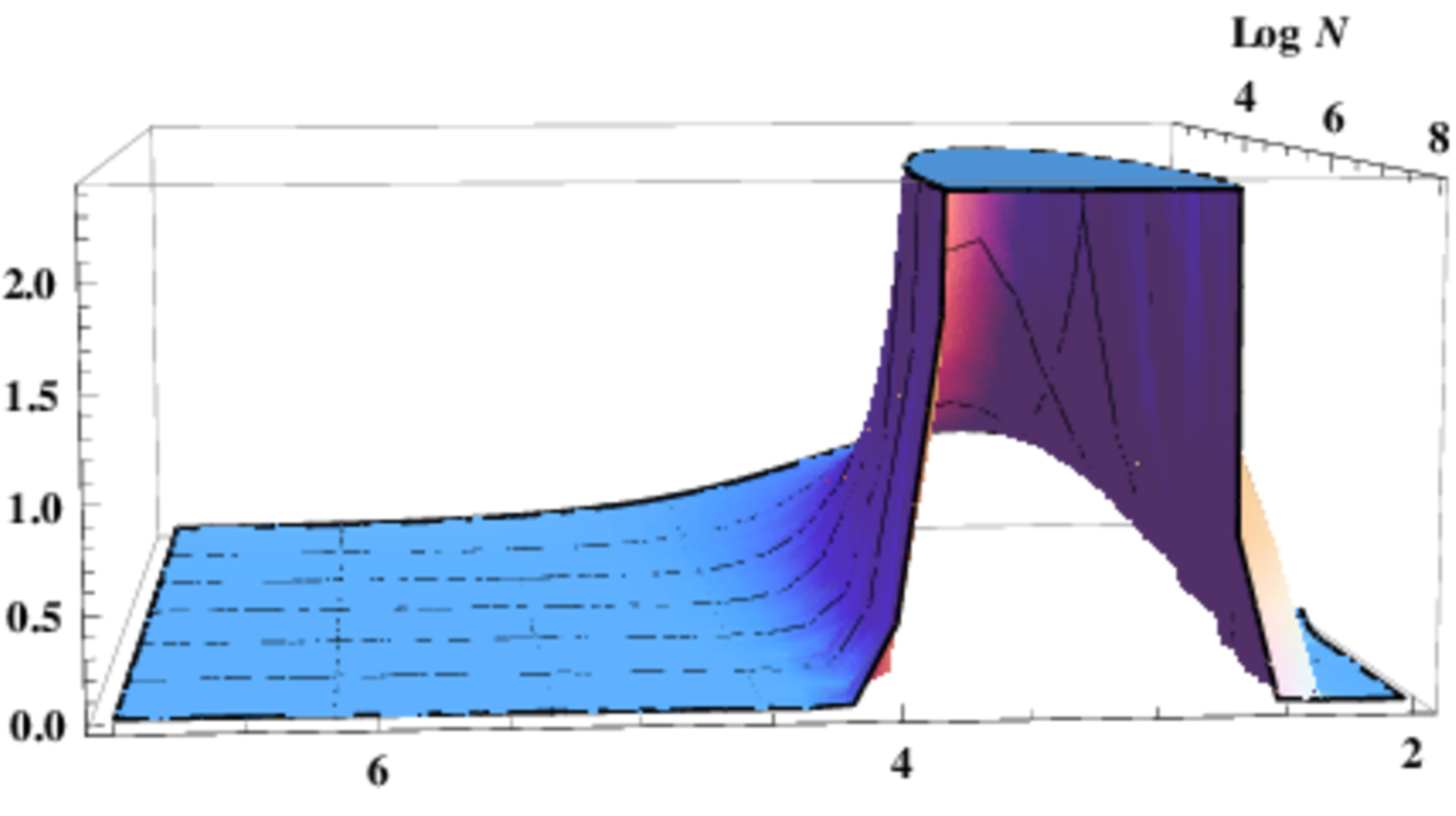}}
{\includegraphics[height=1.65in]{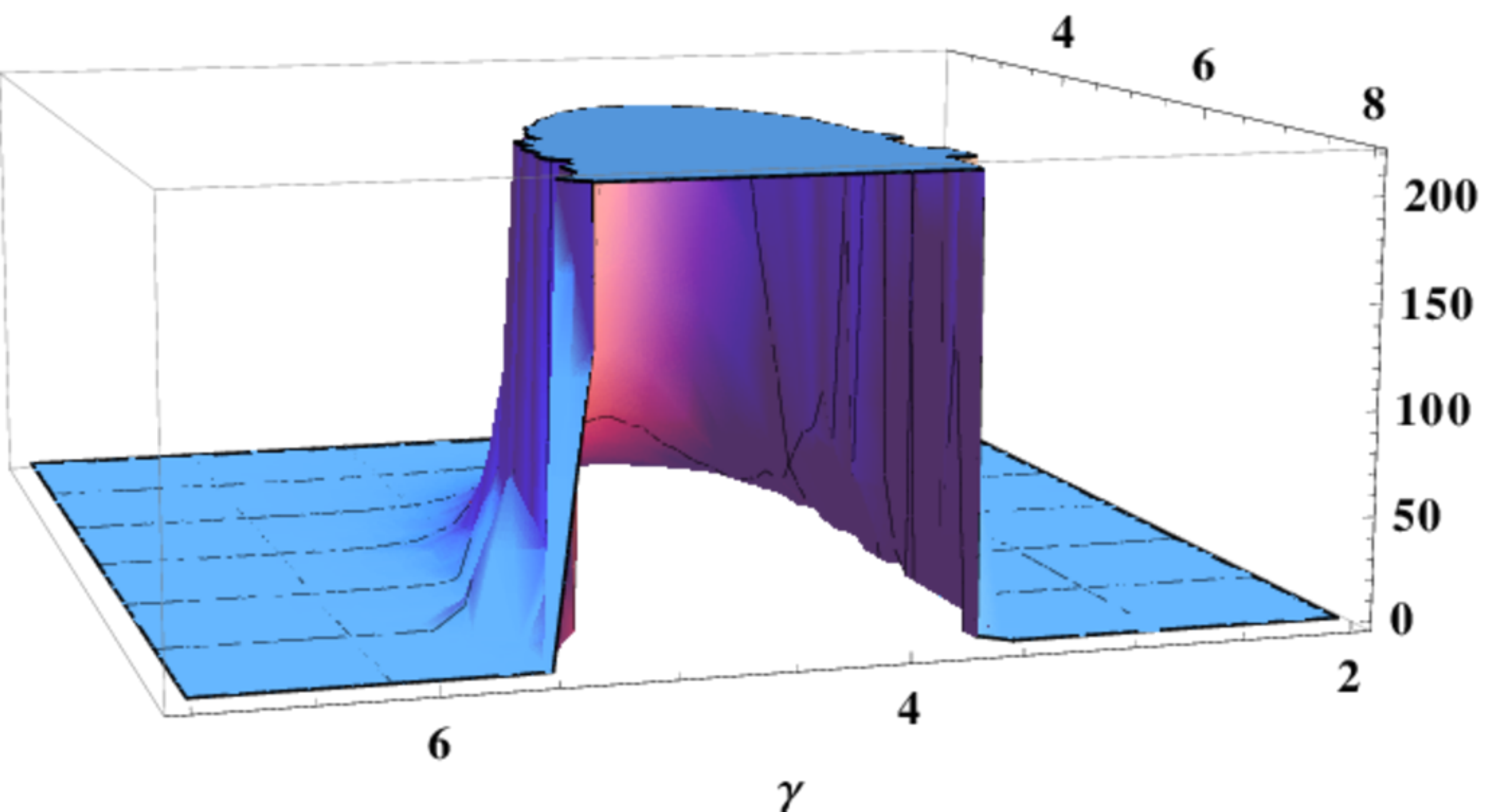}}
\caption{$R_\mathsmaller{\Gamma_3}$ (top with $\gamma_1\simeq 2.5,\gamma_2\simeq 4$) and $R_\mathsmaller{\Gamma_{4c}}$ (bottom with $\gamma_1\simeq 3.5,\gamma_2\simeq 5.7$) as functions of $N$ and $\gamma$,
from numerical integrations of Eq. (\ref{Rclique}) using $10^7$ points per integral. 
\label{fig4.5}
}
\end{figure}

\textit{Singular terms.}
The main message of this Letter is that, 
if the maximal degree $k_{\mathrm{max}}$ of the motif $\Gamma$ is greater than 1, there are contributions which 
make $R_\mathsmaller{\Gamma}$ divergent for $N\to\infty$. Let us analyze the Green function $\left[\Gamma_{kc\times 2}\right]$ which appears in Eq. (\ref{Rclique}).
From Eqs. (\ref{CM})-(\ref{rho}), by enumerating the $2k-1$ nodes of $\Gamma_{kc\times 2}$ with $1,\ldots,2k-1$, $2k-1$ being the central node, we have
\begin{eqnarray}
\label{Gkcx2}
&& \left[\Gamma_{kc\times 2}\right]=\frac{a^{2k-1}}{k^{4\binom{k}{2}}_s}\int \prod_{i=1}^{2k-1} dh_i \prod_{i=1}^{2k-2} h_i^{k-1-\gamma}h_{2k-1}^{2k-2-\gamma} \nonumber \\
&\times& \prod_{i=1}^{2k-2}\left(1+\frac{h_{2k-1}h_i}{k_s^2}\right)^{-1}\prod_{i<j:~i,j\neq 2k-1}\left(1+\frac{h_ih_j}{k_s^2}\right)^{-1} \nonumber.
\end{eqnarray}
We see that $h_{2k-1}$ plays a ``privileged'' role with respect to the other variables $h_1,\ldots,h_{2k-2}$.
When $h_1,\ldots,h_{2k-2}\ll k_s$,
all the factors in the second row of this Eq. remain finite, 
independently of the value of $h_{2k-1}$, which must be integrated from $h_{\mathrm{min}}$ to $h_{\mathrm{max}}=N$.
Such a region of integration, when $(k-1)<\gamma<2(k-1)$, gives the leading contribution to $\left[\Gamma_{kc\times 2}\right]$, and
taking into account that $\left[\Gamma_{kc}\right]=\mathop{O}(1/k_s^{2\binom{k}{2}})$,
the net result is $R_\mathsmaller{\Gamma_{kc}}=\mathop{O}(N^{2k-2-\gamma})$. 
This preliminary analysis is approximate (the actual exponent is smaller) but consistent, and shows a general mechanism that does not
instead apply for $h_{\mathrm{max}}=N^{1/(\gamma-1)}$.
It holds for any motif $\Gamma$, both sparse (\textit{e.g.} cliques) or dense 
(\textit{e.g.} open chains of links): 
$\lim_{N\to\infty}R_\mathsmaller{\Gamma}=\infty$ for $\gamma\in (\gamma_1,\gamma_2)$, where $\gamma_1\approx \mathrm{max\{2,k_{\mathrm{min}}\}}$, and $\gamma_2\approx 2k_{\mathrm{max}}$,
$k_{\mathrm{min}}$ and $k_{\mathrm{max}}$ being the smallest and the largest degrees of $\Gamma$.  
Numerical integrations confirm this phase-transition scenario 
when $h_{\mathrm{max}}\sim N$, while we do not see any divergent behavior for $h_{\mathrm{max}}\sim N^{1/(\gamma-1)}$ (compatibly with Refs. \cite{Bianconi1,Bianconi2,BogunaC}). 
In Fig.~\ref{fig4.5} we show the $k$-clique cases $k=3$ and $k=4$.
This general result is expected to hold for any probability $p(h,h')$ as a function of $hh'/k^2_s$ when $h_{\mathrm{max}}\sim N$. 
An urgent question concerns the possibility to single out ensembles where the non self-averaging terms in $R_\mathsmaller{\Gamma}$ are absent.
Related results will be given elsewhere. 

\section{Conclusions}
Hidden variables models provide a powerful tool to investigate 
complex networks analytically. 
{
In this framework we present a first systematic
analysis of the  fluctuations of the density of  motifs.
Surprisingly, if the cut-off $h_{\mathrm{max}}$ of the model is properly chosen
as to reproduce a power-law distribution $P(k)\propto k^{-\gamma}$ having support $k\leq N$, \textit{i.e.}, $h_{\mathrm{max}}=\mathop{O}(N^\lambda)$, with $\lambda\geq 1$, 
a phase transition scenario emerges,
with self-averaging regions separated from  non-self-averaging regions~\footnote{
{We postpone the issue of the minimal value of $\lambda$, $\lambda_c$, above which such a picture still applies. However, for $\gamma>2$, $1 \geq\lambda_c>1/(\gamma-1)$.}
}.
Under the simple null-model assumption that the system obeys a power-law, the potential practical consequences of such a picture are dramatic.
The existence of non-self-averaging regions implies the instability of the substructures composing 
a network to even small perturbations: patterns and communities
structures observed in a  given network-realization,  will have  a totally
different configuration in another network-realization. 
For the same reason, in simulating  complex  networks,  the  presence  of  large  fluctuations
implies a very demanding statistics in order to have a fair evaluation
of the averages of the observables of interest: given a motif $\Gamma$, if we are in a non-self-averaging region,
the number of samples that we need to properly evaluate the average $\mediaT{n_\Gamma}$ 
will be a growing function of the system size $N$. Hence, since the evaluation of
$n_\Gamma$ already requires $N$ operations per each sample, the total number of operation to evaluate $\mediaT{n_\Gamma}$ scales as $N^{1+\alpha}$, with $\alpha>0$.

Our analysis could also explain why most of the observed networks have a small value of $\gamma$ \cite{NewData}. 
In fact, with such small values, even small motifs and communities are guaranteed to be stable as soon as they have $k_{\mathrm{min}}\gtrsim \gamma$.  
Whereas, for $\gamma$ larger, communities for which $k_{\mathrm{min}}\lesssim\gamma\lesssim 2k_{\mathrm{max}}$ will be unstable to small perturbations.
In practical terms this means that, between  two networks, one with say,  
$\gamma=2.3$, and another with  say, $\gamma=2.1$,  the  latter is  
more stable or, in other words,  it has a larger probability to exist.   
Moreover, our analysis  also  shows  that 
fluctuations  of  motifs  become  negligible  for  $\gamma\to 2$. 
This latter observation is consistent with the fact that 
the  entropy of networks  for  $\gamma\to  2$  goes to  zero \cite{Bianconi3,Genio}.} 

We have specialized the analysis to the hidden-variable
model defined by Eqs. (\ref{CM})-(\ref{rho}). 
A similar self- \textit{vs} non-self-averaging phase transition scenario is expected to hold
for any hidden-variable model characterized by power laws. 

\begin{acknowledgments}
Work supported by DARPA grant No.\ HR0011-12-1-0012; NSF grants No.\ CNS-0964236 and CNS-1039646; and by Cisco Systems.
We thank D. Krioukov from whom this research was inspired,
and G. Bianconi, Z. Toroczkai, M. Bogu$\mathrm{\tilde{n}}$\'a, and C. Orsini for useful discussions.
\end{acknowledgments}

\end{document}